\documentclass[preprint,review, 3p, 12pt]{elsarticle}

\usepackage{amssymb}
\usepackage{amsmath}
\usepackage{xcolor}
\biboptions{sort&compress}

\journal{Chaos, Solitons and Fractals}

\begin{document}

\begin{frontmatter}



\title{Analysis of persistent and antipersistent time series with the Visibility Graph method}

\author[inst1]{Macarena Cádiz\corref{cor1}}
\ead{macarena.cadiz@ug.uchile.cl}
\author[inst1]{Iván Gallo-Méndez\corref{cor1}}\ead{ivan.gallo@ug.uchile.cl}
\author[inst1]{Pablo S. Moya\corref{cor1}}\ead{pablo.moya@uchile.cl}
\author[inst1]{Denisse Pastén\corref{cor1}}\ead{denisse.pasten.g@gmail.com}
\cortext[cor1]{Corresponding authors}
\affiliation[inst1]{organization={Departamento de Física, Facultad de Ciencias, Universidad de Chile},
            addressline={Las Palmeras 3425}, 
            city={Santiago},
            postcode={7800003}, 
            country={Chile}}

%%%%%%%%%%%%%%%%%%%%%%%%%%%%%%%%%%%%%%%%%%%%%%%%%%%%%%%%%%%%%%%%%%%%%%%%%%%%%%%%%%%%%%%
%Abstract
%%%%%%%%%%%%%%%%%%%%%%%%%%%%%%%%%%%%%%%%%%%%%%%%%%%%%%%%%%%%%%%%%%%%%%%%%%%%%%%%%%%%%%%
\begin{abstract}

In this work, we investigate a range of time series, including Gaussian noises (white, pink, and blue), stochastic processes (Ornstein–Uhlenbeck, fractional Brownian motion, and Lévy flights), and chaotic systems (the logistic map), using the Visibility Graph (VG) method. We focus on the minimum number of data to use VG and on two key descriptors: the degree distribution $P(k)$, which often follows a power-law $P(k) \sim k^{-\gamma}$, and the Hurst exponent $H$, which identifies persistent and antipersistent time series. While the VG method has attracted growing attention in recent years, its ability to consistently characterize time series from diverse dynamical systems remains unclear. Our analysis shows that the reliable application of the VG method requires a minimum of 1000 data points. Furthermore, we find that for time series with a Hurst exponent $H \leq 0.5$, the corresponding critical exponent satisfies $\gamma \geq 2$. These results clarify the sensitivity of the VG method and provide practical guidelines for its application in the analysis of stochastic and chaotic time series.

\end{abstract}

%%%%%%%%%%%%%%%%%%%%%%%%%%%%%%%%%%%%%%%%%%%%%%%%%%%%%%%%%%%%%%%%%%%%%%%%%%%%%%%%%%%%%%%
% Graphical Abstract
%%%%%%%%%%%%%%%%%%%%%%%%%%%%%%%%%%%%%%%%%%%%%%%%%%%%%%%%%%%%%%%%%%%%%%%%%%%%%%%%%%%%%%%
\begin{graphicalabstract}
    \begin{figure}[h] 
        \includegraphics[width=0.85\linewidth]{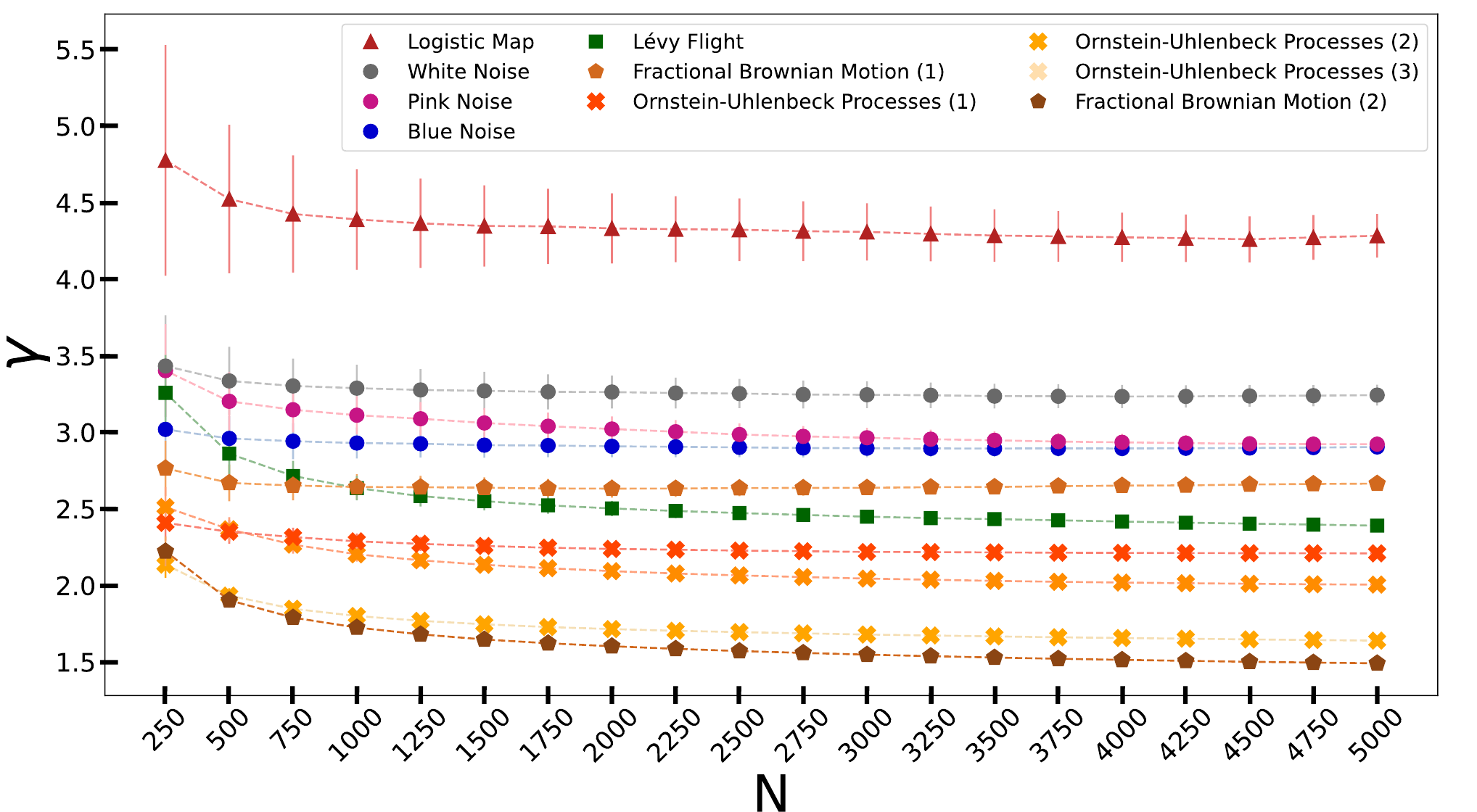}
    \end{figure}
    \begin{figure}[h] 
        \vspace{-0.5cm}
        \hspace{-0.3cm}
        \includegraphics[width=0.7\linewidth]{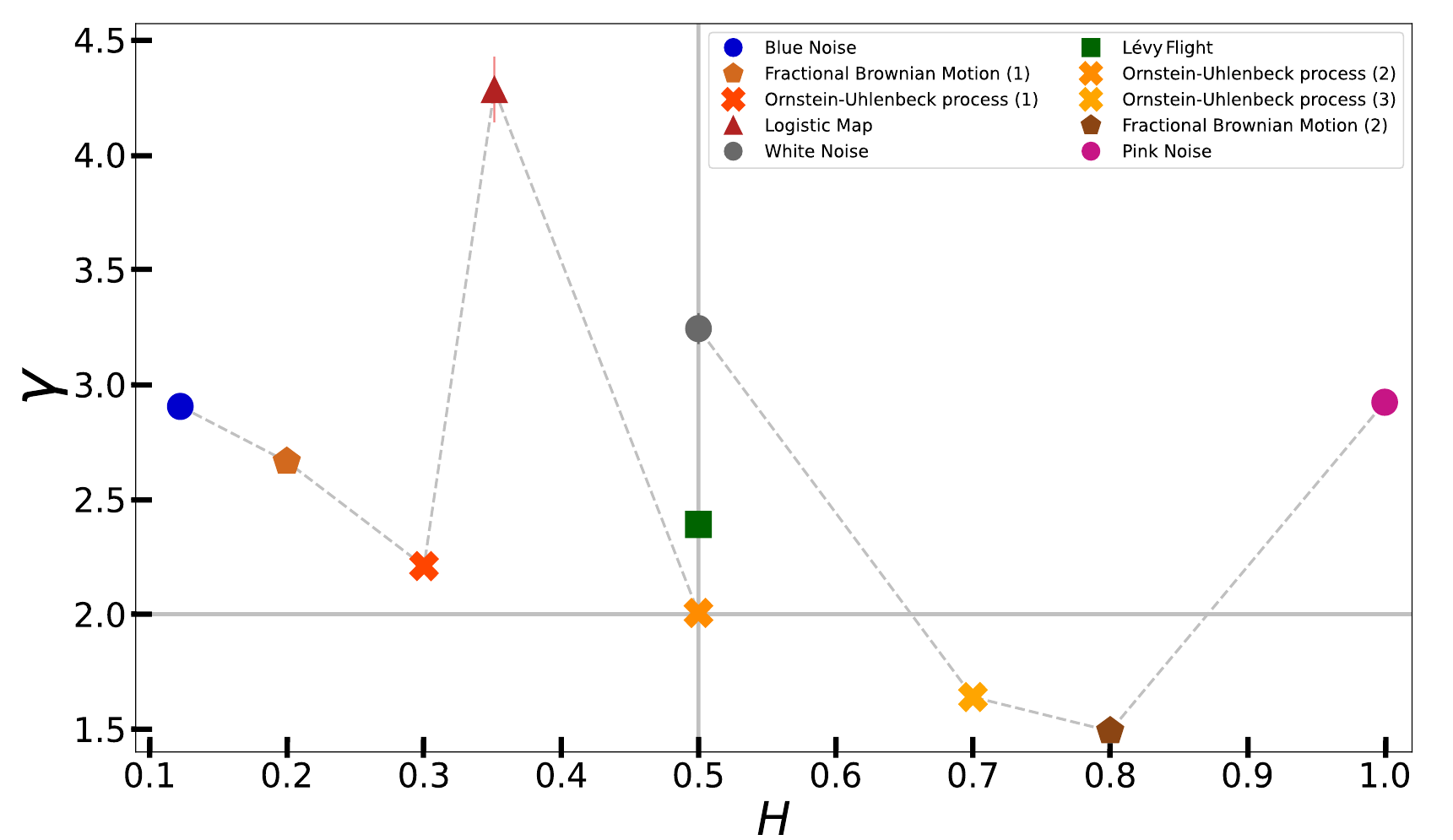}
    \end{figure}
\end{graphicalabstract}

\begin{highlights}
    \item For consistent and confident use of the Visibility Graph method, a minimum of 1000 data points is recommended.
    \item When a time series has a Hurst exponent $H \leq 0.5$, the corresponding degree distribution consistently exhibit a critical exponent $\gamma \geq 2$.
\end{highlights}

%%%%%%%%%%%%%%%%%%%%%%%%%%%%%%%%%%%%%%%%%%%%%%%%%%%%%%%%%%%%%%%%%%%%%%%%%%%%%%%%%%%%%%%
% Keywords
%%%%%%%%%%%%%%%%%%%%%%%%%%%%%%%%%%%%%%%%%%%%%%%%%%%%%%%%%%%%%%%%%%%%%%%%%%%%%%%%%%%%%%%
\begin{keyword}
Visibility Graph \sep Critical exponent \sep Hurst exponent \sep Power-law 
\end{keyword}

\end{frontmatter}



%%%%%%%%%%%%%%%%%%%%%%%%%%%%%%%%%%%%%%%%%%%%%%%%%%%%%%%%%%%%%%%%%%%%%%%%%%%%%%%%%%%%%%%
% Introduction
%%%%%%%%%%%%%%%%%%%%%%%%%%%%%%%%%%%%%%%%%%%%%%%%%%%%%%%%%%%%%%%%%%%%%%%%%%%%%%%%%%%%%%%
\section{Introduction}

In graph theory, individual entities are referred to as nodes (or vertices), and the relationships or interactions between them are represented by edges (or links). The figure resulting from this method is known as a graph or, in more applied terms, a network. Some particularly interesting types of graphs in the study of real systems are complex networks. Complex networks relate to emergent phenomena, that is, properties that are not visible through individual elements of the system. These phenomena are visible in the complex network through the use of metrics and network topology. This type of network has proven to be a useful tool for studying a wide variety of systems in nature and society~\cite{albert2002statistical}, spanning areas such as computational science, food webs, cell behavior, social networks, wildfires, and even earthquakes~\cite{anderson1993cayley, girvan2002community, li2014symmetric, montoya2002small, gallo2025time, telesca2020analysis}.

Complex networks have proven to be powerful tools for characterizing time series. In 2008, Lacasa et al.~\cite{lacasa2008time} proposed a computational algorithm for complex networks known as the Visibility Graph (VG), which maps the time series into nodes of a complex network using a geometric criterion based on the magnitude of the data and its visibility relative to other data in the series. Also, Lacasa et al.~\cite{lacasa2009visibility} obtained a relation that allows one to obtain the Hurst exponent for fractional Brownian motion using the visibility graph method. A year later, Luque et al.~\cite{luque2009horizontal} proposed the Horizontal Visibility Graph (HVG) algorithm, which is a more restrictive version of VG whose criterion depends on the horizontal visibility of the time series and the temporal direction. The interesting aspect of HVG is that, in conjunction with the Kullback-Leibler Divergence (KLD), it also facilitates the study of time series reversibility~\cite {lacasa2012time}. The visibility graph algorithms have proven to be very useful to characterize time series of physical systems. For instance, Suyal et al.~\cite{suyal2014visibility} and Acosta-Tripailao et al.~\cite{acosta2023AA}, using the HVG algorithm, show that the degree of irreversibility ($D$) varied systematically with the solar cycle, indicating that solar wind fluctuations become more or less reversible depending on the solar activity phase. Zhang et al.~\cite{zhang2021analysis}, using the VG algorithm, show that Xi'an's temperature dynamics are scale-invariant, robust over time, and exhibit seasonal decoupling, especially in summer, which aligns with a propensity for extreme weather. Acosta-Tripailao et al.~\cite{acosta2022assigning}, using the HVG algorithm, characterized blazar light curves by assigning degrees of stochasticity. They found chaotic behavior in the time series for the BL Lac sources and correlated stochastic behavior in the time series for the FSRQ sources. Additional studies have successfully applied visibility graph techniques to domains such as solar activity, river flow fluctuations, structural biology, non-Maxwellian turbulent plasmas, wall turbulence, pulsating stars, solar wind, sandpiles, and others~\cite{zurita2023characterizing,braga2016characterization,baoli2023,saldivia2024using,iacobello2018visibility,munoz2021analysis,acosta2021applying,Adami_2025,Yu_2016,xie_2011,masoomy2023PhysRevE}.

On the other hand, a useful parameter for characterizing time series is the Hurst exponent $H$, which quantifies the degree of long-term memory or temporal persistence in the data~\cite{hurst1951long}. The Hurst exponent has proven valuable across a wide range of disciplines. In physiology, Lebamovski and Gospodinova~\cite{lebamovski2025investigating} applied Detrended Fluctuation Analysis (DFA) \cite{peng1994mosaic} to heart rate variability signals recorded during a virtual reality task designed to induce cognitive stress. They estimated the Hurst exponent from the scaling behavior of the fluctuation function to quantify long-range temporal correlations in cardiac dynamics, showing that stress reduces the persistence of heart rate fluctuations. In neuroscience, Dong et al.~\cite{dong2018hurst} showed that the Hurst exponent provides a nuanced fractal marker of healthy brain aging. Other notable applications include traffic modeling, hydrology, and machine learning, among others \cite{chand2021modeling, movahed2008fractal, raubitzek2023scaling}. The relationship between the Hurst exponent and the exponent associated with the degree distribution of the resulting network has been analytically established only for the Horizontal Visibility Graph algorithm, as shown by Luque et al.~\cite{luque2009horizontal}. In contrast, the potential of the standard Visibility Graph algorithm to characterize long-term correlations remains relatively underexplored.

We analyze diverse time series using the standard Visibility Graph algorithm, knowing that it benefits systems outside of thermodynamic equilibrium. We use several time series from deterministic to stochastic, Markovian, and non-Markovian, where the former exhibits short-range correlations (short memory), meaning the future depends only on the immediate past. At the same time, the latter shows long-range correlations and typically corresponds to non-equilibrium processes~\cite{vacchini2011markovianity}. One of the key points of this study is to determine the minimum number of data points needed to use the VG and obtain reliable results. This result is of great interest when working with data that have been measured and the number of data points is only a few thousand, as is the case in studies of astrophysical systems, such as blazars or pulsars, among others.

Here we explore in depth the limits of VG in terms of the amount of data and associate the exponents of the degree distribution of the resulting graph with a classic exponent for nonlinear systems, such as the Hurst exponent, aiming to explore how memory and persistence properties influence the degree distribution and overall network topology generated by the VG method. 

The structure of this paper is as follows: In Section~\ref{sec: methodology}, we introduce the time series analyzed in this study. Section~\ref{sec: vg} details the Visibility Graph algorithm, while Section~\ref{sec: hurst} discusses the computation and relevance of the Hurst exponent. In Section~\ref{sec: results}, we present and analyze the results, focusing on the relationship between the Visibility Graph metrics and the Hurst exponent. Finally, Section~\ref{sec: conclusions} concludes the paper with a summary of our main findings and their implications.

%%%%%%%%%%%%%%%%%%%%%%%%%%%%%%%%%%%%%%%%%%%%%%%%%%%%%%%%%%%%%%%%%%%%%%%%%%%%%%%%%%%%%%%
% Methodology
%%%%%%%%%%%%%%%%%%%%%%%%%%%%%%%%%%%%%%%%%%%%%%%%%%%%%%%%%%%%%%%%%%%%%%%%%%%%%%%%%%%%%%%
\section{Times Series}
\label{sec: methodology}

In this study, we analyze ten different types of time series artificially generated by their definition:

\begin{itemize}
    \item \textbf{Logistic map}. This is one of the most well-known dynamical systems, which is also a Markovian process \cite{lorenz1964problem, steven2015strogatz}, defined by the equation 
        \begin{equation}
            \label{eq: log_map}
            x_{t+1} =  r x_t (1 - x_t),
        \end{equation}

    \noindent where $0 \leq r \leq 4$ and $t$ are the time steps. It is well known that the logistic map exhibits chaotic behavior as the control parameter $r$ approaches 4. In this work, we focus on the chaotic regime by setting $r = 3.8$, a value that ensures sensitivity to initial conditions and complex dynamics. While the logistic map has been extensively studied, the behavior of complex network metrics derived from its chaotic time series remains relatively unexplored.
    
    \item \textbf{Noise spectra}. These series correspond to a stochastic signal generated by non-Markovian processes~\cite{luczka2005non}. The different ``colors" of noise are characterized by their power spectral density (PSD) per unit bandwidth, which typically follows the form $S(f) \propto 1/f^{\beta}$~\cite{zhivomirov2018method}. This means the PSD is not constant but depends on the frequency~$f$. Specifically, white noise has $\beta = 0$, pink noise $\beta = 1$, and blue noise $\beta = -1$. In this work, we use noise spectra in the range of 1 kHz.

    \item \textbf{Fractional Brownian Motion (fBm)}. The fBm is a generalized form of Brownian Motion, whose increments are stationary but not independent, making it a non-Markovian process~\cite{mandelbrot1968fractional}. Its definition is

    \begin{equation}
        \label{eq: fbm}
        \begin{split}
         B_H(t) = \frac{1}{\Gamma(H+1/2)} \Bigg[\int_{-\infty}^{0} ((t-u)^{H-1/2}
        - (-u)^{H-1/2}) \, dB(u) \\
        + \int_{0}^{t} (t-u)^{H-1/2} \, dB(u) \Bigg],
        \end{split}
    \end{equation}

    \noindent with $t>0$, $H$ is Hurst exponent, and $B(u)$ is Brownian motion. We calculate the time series through the Davies-Harte method~\cite{davies1987tests}. In this work, we use $H=0.2$ and $H=0.8$.

    \item \textbf{Ornstein-Uhlenbeck process}. In terms of fractality, this is a monofractal system characterized by being the only process that is simultaneously Gaussian, Markovian, and stationary~\cite{tabar2019analysis}. It is defined by 

    \begin{equation}
    \label{eq: o-u_process}
        dx_t = -\theta x_t \, dt + \sigma \, dB_t^H,
    \end{equation}

    \noindent with steps $t$, $\theta >0$, $\sigma > 0$, and $B_t^H$ being Brownian motion. For this type of time series, we employ values of $H=0.3$, $H=0.5$, and $H=0.7$.

    \item \textbf{Lévy flights}. As part of our analysis of fractal series, we also include these stochastic processes with stationary, independent increments and power-law distributed step lengths, making them suitable for modeling systems with anomalous diffusion or extreme variability~\cite{applebaum2009levy}. Lévy flights are a type of Markovian stochastic process characterized by jump lengths that follow a heavy-tailed probability distribution (PDF)~\cite{chechkin2008introduction}. Specifically, the probability density function $\lambda(x)$ decays for large $x$ as 

    \begin{equation}
        \label{eq: levy_flight}
        \lambda(x) \approx |x|^{(-1-\alpha)},
    \end{equation}

    \noindent with $0 < \alpha < 2$, being the stability index of the distribution. In this work, we use $\alpha=1.3$ because it gives a Hurst exponent $H = 0.5$ as computed in~\cite{masoomy2023PhysRevE}.
    
\end{itemize}

%%%%%%%%%%%%%%%%%%%%%%%%%%%%%%%%%%%%%%%%%%%%%%%%%%%%%%%%%%%%%%%%%%%%%%%%%%%%%%%%%%%%%%%
% Visibility Graph (VG)
%%%%%%%%%%%%%%%%%%%%%%%%%%%%%%%%%%%%%%%%%%%%%%%%%%%%%%%%%%%%%%%%%%%%%%%%%%%%%%%%%%%%%%%
\section{Visibility Graph}
\label{sec: vg}

\begin{figure}[h]
    \centering
    \includegraphics[width=0.8\linewidth]{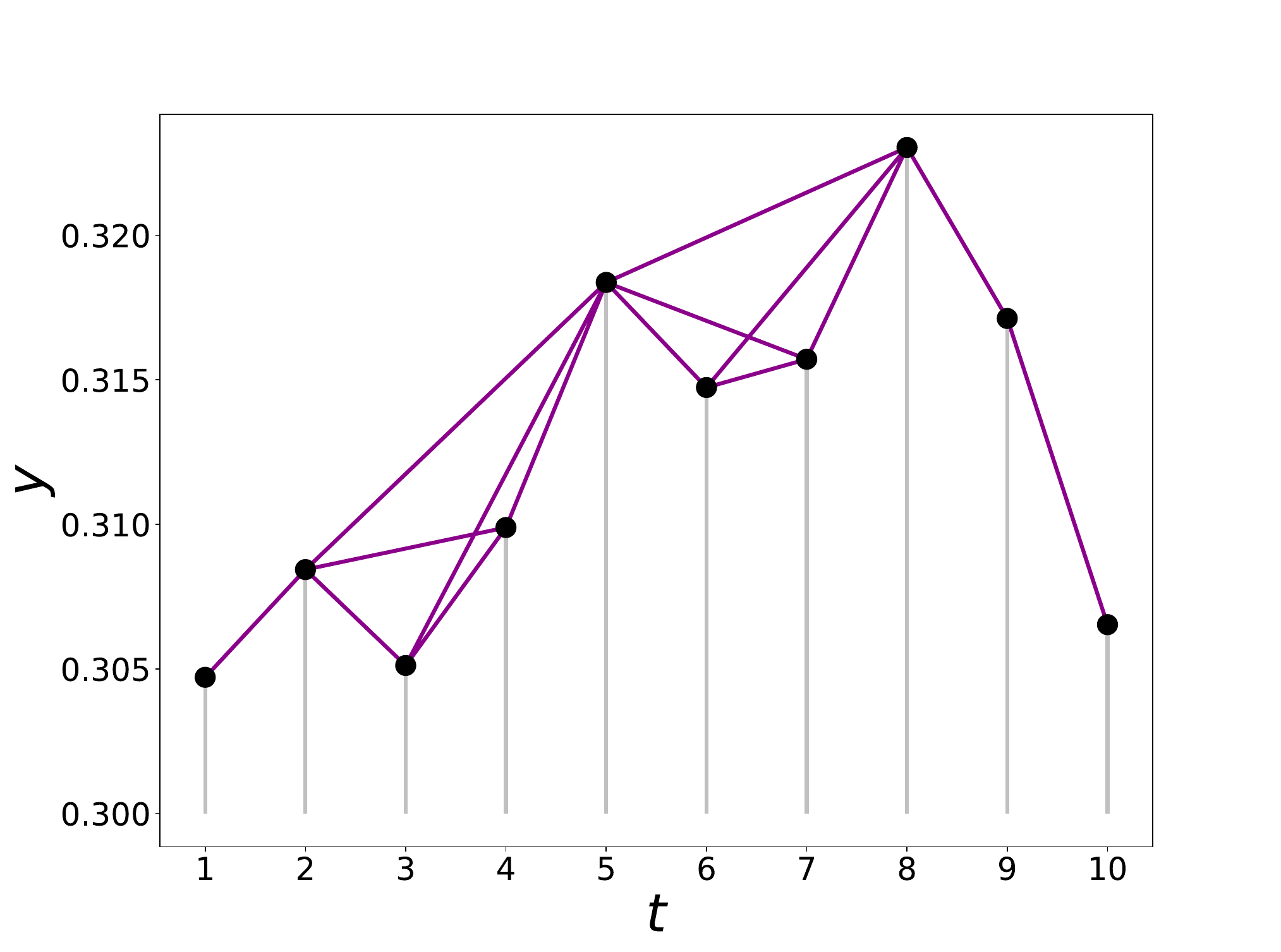}
    \caption{Illustration of the VG algorithm applied to the last 10 data points from a 5000-point fractional Brownian motion time series. The data points (nodes) are represented as black circles, with vertical lines indicating their heights. Purple lines depict the visibility-based connections (links) formed between the nodes.}
    \label{fig:vg_example}
\end{figure}

The Visibility Graph algorithm (VG) maps data from a time series to nodes of a complex network, where these nodes can be connected through links~\cite{lacasa2008time}. The condition under which a node is established to be connected to another is a ``visibility condition'', and this is based on a geometric criterion. For two nodes to have visibility, it must be possible to draw a straight line between them without any intermediate node intersecting it, as we can see in Fig. \ref{fig:vg_example}. 

More formally, two arbitrary data points $(t_a, y_a)$ and $(t_b, y_b)$ will have visibility and consequently will become two connected nodes of the associated graph if any other data point $(t_c, y_c)$ between them fulfills:

\begin{equation}
    \label{eq: visibility_condition}
     y_c < y_b + (y_a - y_b) \frac{t_b - t_c}{t_b - t_a}.
\end{equation}

From the presented characteristics in Fig. \ref{fig:vg_example}, we can notice that the graph produced by the algorithm will always be connected, undirected, and invariant under transformations of the data in the series.

\begin{figure}[hp]
    \centering
    \includegraphics[width=0.75\linewidth]{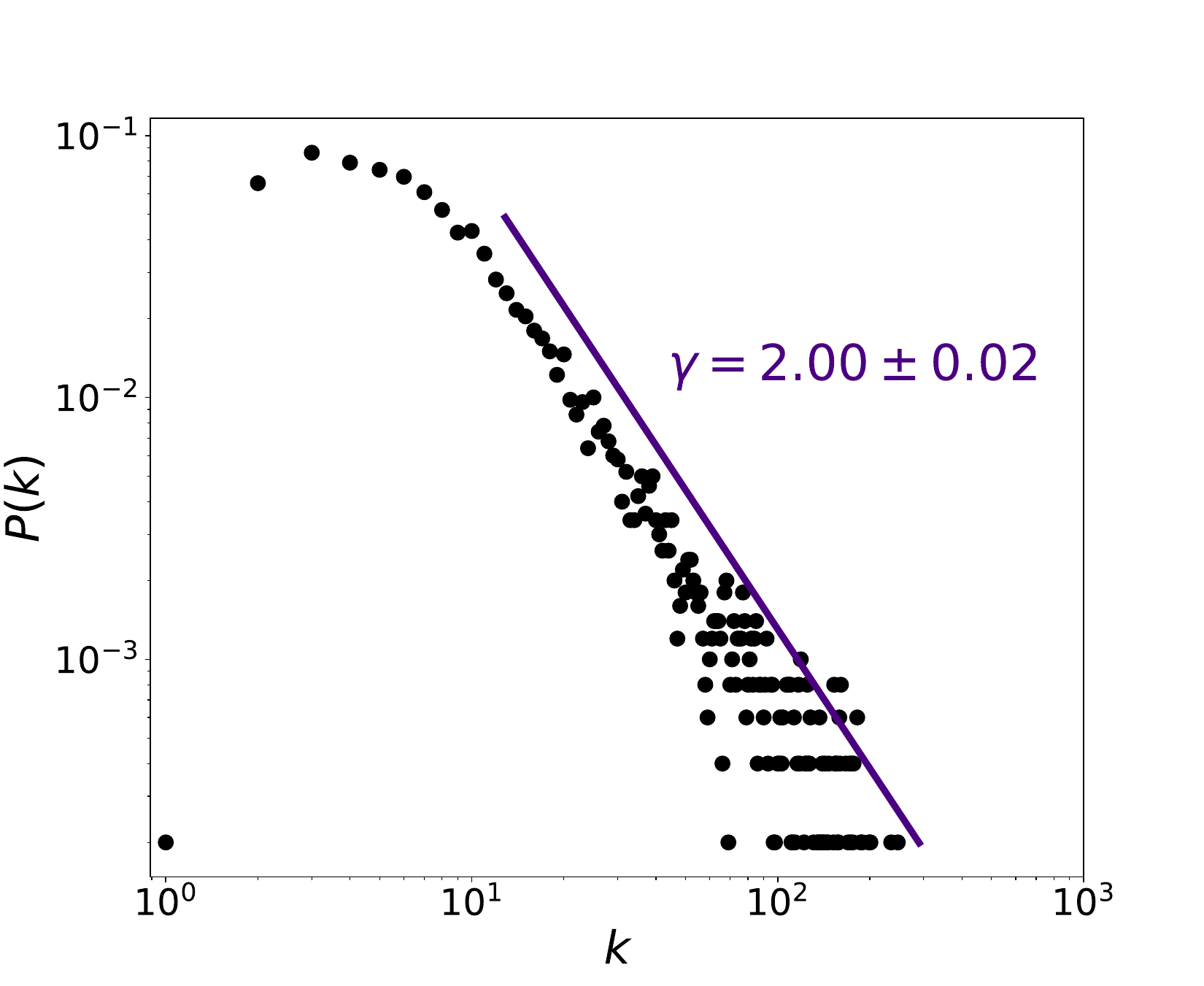}
    \caption{Degree distribution of the associated visibility graph created with the complete series of fractal Brownian motion of 5000 data values (for the graph, see Fig. \ref{fig:vg_example}), where the critical exponent of the power-law is $\gamma =  1.99 \pm 0.02$, with $k_{\text{min}} = 7$.}
    \label{fig:pk_example}
\end{figure}

A fundamental metric for characterizing a graph is the degree distribution $P(k)$, which represents the probability of finding a node with degree $k$ in the network, where $k$ is the number of connections a node has. It is known that most real systems are scale-free~\cite{barabasi1999emergence}, meaning their degree distribution behaves as a power-law in the form

\begin{equation}
    \label{eq: power_law}
        P(k) \sim k^{-\gamma},
\end{equation}

\noindent where $\gamma$ is known as the critical exponent in terms of the connectivity in the scale-free system~\cite{albert2002statistical}. For example, we can see the degree distribution of a fractional Brownian motion series of 5000 data values in Fig. \ref{fig:pk_example}, which behaves as a power-law. This result is also consistent with the proposal of Lacasa et al.~\cite{lacasa2008time}, who calculated the critical exponent for a fractional Brownian motion of $10^6$ data points.

The $\gamma$ exponent can be obtained through a linear regression with the Maximum Likelihood Estimation (MLE) method~\cite{telesca2012analysis}:

\begin{equation}
    \label{eq: gamma_mle}
    \gamma = 1 + n \left[\sum_i^n \ln \left( \frac{k_i}{k_{\text{min} - 0.5}} \right) \right]^{-1},
\end{equation}

\noindent where $k_{\text{min}}$ is the degree where the distribution starts to behave like a power-law and $n$ is the number of values with $k \geq k_{\text{min}}$. Furthermore, the error associated with the $\gamma$ value can be obtained with the standard deviation:

\begin{equation}
    \label{eq: error_mle}
    \delta \gamma = \frac{\gamma - 1}{\sqrt{n}}.
\end{equation}

It is important to note that in this context, we use MLE instead of linear fit-based methods because it is proven that the MLE method is more robust for estimating the exponent of power-laws~\cite{goldstein2004problems}.

To assess the reliability of the power-law fits, we employ the Kolmogorov–Smirnov (KS) test~\cite{clauset2009power}. The test compares cumulative distribution functions (CDFs): the empirical CDF, denoted as $S(k)$, obtained from the degree data with $k \geq k_{\min}$, and the theoretical CDF, denoted as $F(k)$, corresponding to the fitted power-law distribution with exponent $\alpha$ and the same lower bound $k_{\min}$. The KS distance $D$ is then defined as

\begin{equation}
    D = \max_{k \geq k_{\min}} |S(k) - F(k)|, 
\end{equation}

\noindent which represents the maximum vertical distance between the empirical distribution and the fitted power-law model. The value of $D$ lies in the interval $[0, 1]$, where smaller values of~$D$ indicate a closer agreement between data and model, whereas larger values signal weaker fits.

%%%%%%%%%%%%%%%%%%%%%%%%%%%%%%%%%%%%%%%%%%%%%%%%%%%%%%%%%%%%%%%%%%%%%%%%%%%%%%%%%%%%%%%
% Hurst Exponent
%%%%%%%%%%%%%%%%%%%%%%%%%%%%%%%%%%%%%%%%%%%%%%%%%%%%%%%%%%%%%%%%%%%%%%%%%%%%%%%%%%%%%%%
\section{Hurst Exponent}
\label{sec: hurst}

The Hurst exponent is a valuable tool for characterizing the dynamics of a time series~\cite{hurst1951long}. It quantifies the presence of long-range dependence and self-similarity in the data. The exponent $H$ takes values in the range $[0,1]$, and its value provides insight into the underlying correlation structure:

\begin{itemize}
    \item $H < 0.5$: The time series is said to be antipersistent, meaning that if it increases during one period, it will likely decrease in the following period, and vice versa.
    \item $H=0.5$: The time series behaves like an uncorrelated or random process.
    \item $H>0.5$: The time series exhibits long-term memory, implying that past events positively influence future events.
\end{itemize}

To estimate the Hurst exponent of a time series, we use the Detrended Fluctuation Analysis (DFA).

\subsection{Detrended Fluctuation Analysis}

DFA is a method designed to quantify long-range correlations in a time series \cite{peng1994mosaic, peng1995quantification, kantelhardt2001detecting, rogers2022fractal}. Considering a time series $x_k$ with $N$ elements, and its mean value $\langle x \rangle$, the first step is to construct the profile of the time series:

\begin{equation*}
    Y(i) \equiv \sum_{k=1}^{i}[x_k - \langle x \rangle], \quad i=1, \dots, N.
\end{equation*}

This profile $Y(i)$ is divided into $N_s \equiv \text{int}(N/s)$ non-overlapping segments of equal length~$s$. Since $N$ is not always a multiple of $s$, the same division is performed starting from the opposite end of the series, resulting in a total of $2N_s$ segments. 

For each segment $\nu$, we determine the local trend $y_\nu(i)$ by performing a least-squares fit. Then, we calculate the variance for each segment:

\begin{equation*}
    F^2(\nu, s) \equiv \frac{1}{s} \sum_{i=1}^{s} \{ Y[(\nu-1)s + i] -y_\nu (i) \}^2.
\end{equation*}

Finally, we average over all segments to obtain the fluctuation function:

\begin{equation*}
    F(s) \equiv \left(\frac{1}{2N_s} \sum_{\nu=1}^{2N_s} F^2(\nu, s) \right)^{1/2}.
\end{equation*}

The fluctuation function $F(s)$ typically exhibits a scaling behavior with segment size $s$:

\begin{equation*}
    F(s) \sim s^{H},
\end{equation*}

\noindent where $H$ is the Hurst exponent.

%%%%%%%%%%%%%%%%%%%%%%%%%%%%%%%%%%%%%%%%%%%%%%%%%%%%%%%%%%%%%%%%%%%%%%%%%%%%%%%%%%%%%%%
% Results
%%%%%%%%%%%%%%%%%%%%%%%%%%%%%%%%%%%%%%%%%%%%%%%%%%%%%%%%%%%%%%%%%%%%%%%%%%%%%%%%%%%%%%%
\newpage
\section{Results}
\label{sec: results}

%%%%%%%%%%%%%%%%%%%%%%%%%%%%%%%%%%%%%%%%%%%%%%%%%%%%%%%%%%%%%%%%%%%%%%%%%%%%%%%%%%%%%%%
\subsection{Sensitivity on the number of data}

Complex networks have proven to be highly effective for characterizing time series, particularly in contexts where data availability is limited. However, the notion of what constitutes ``limited'' or ``few'' data remains somewhat ambiguous. To our knowledge, a systematic study assessing the impact of time series length on the resulting network properties, such as the degree distribution, has not yet been carried out. To address this gap, we analyze the degree distributions of various types of time series, beginning with sequences of 5000 data points.
To evaluate the sensitivity of the VG method to data availability, we analyze how the estimated power-law exponent $\gamma$ of the degree distribution varies as the time series is progressively shortened. Specifically, we randomly remove 250 data points at a time, preserving the order of the points in the time series, taking advantage of the fact that visibility graphs are insensitive to the presence of gaps and the order of removal. This process is repeated iteratively until only 250 points remain.

\begin{figure}[hp]
    \centering
    \includegraphics[width=1.\linewidth]{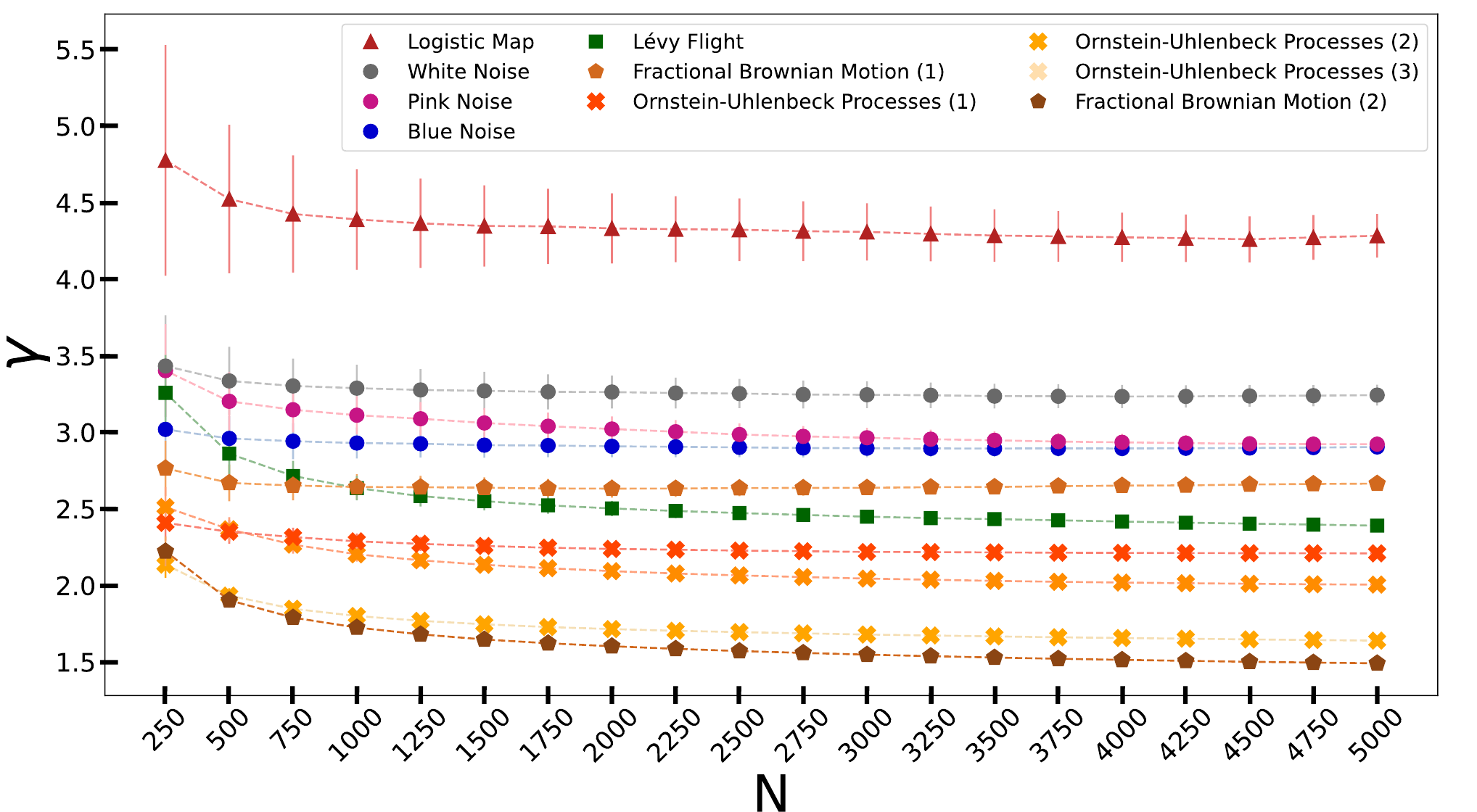}
    \caption{Variation of the $\gamma$ as a function of the time series length $N$. The markers indicate the estimated $\gamma$ values for each time series, while the vertical lines represent the corresponding error bars, where in some cases (such as Ornstein-Uhlenbeck processes), the error bars are not visible due to their very small magnitude.} 
    \label{fig: vg_validation}
\end{figure}

For each time series, we perform this random reduction procedure 1000 times, calculating the corresponding $\gamma$ value for each subseries. The results are then averaged to obtain a robust estimate of how $\gamma$ evolves with decreasing series length. This approach enables us to determine the minimum number of data points necessary for the critical exponent $\gamma$ to stabilize. This analysis is shown in Fig.~\ref{fig: vg_validation}. We can conclude that a reliable value, such that the results of applying the algorithm are robust, is 1000 or higher, as can be seen in Table~\ref{tab: gammas_lengths}. Regardless, with careful consideration, 500 is still a practical limit. In Table~\ref{tab: gammas_lengths} we show the results for data sets from 250 data points to 100\,000 data points to show the robustness of the method. In addition, we have applied a Kolmogorov-Smirnov test and computed the distance $D$ to examine the goodness-of-fit of the power-law obtained from the different data sets. The results of this test are shown in Table~\ref{tab: KS_distance}. We can observe that the value of the distance $D$ decreases for all time series as the number of data points increases, except for the logistic map, a difference that will be analyzed in Section~\ref{tsmenorH}.

\begin{table}[hp]
    \centering
    \begin{tabular}{|c|c|c|c|c|}
        \hline
        \textbf{Time Series} &$\gamma \, (250)$ & $\gamma \, (10^3)$& $\gamma \, (5 \times 10^3)$ & $\gamma \, (10^5)$ \\
        \hline
        Blue Noise  & $2.64 \pm 0.14$ & $2.92 \pm 0.11$ & $2.90 \pm 0.05$ & $3.03 \pm 0.01$  \\
        \hline
        Fractional Brownian Motion (1) & $2.45 \pm 0.11$ & $2.69 \pm 0.09$ & $2.66 \pm 0.04$ & $2.70 \pm 0.01$  \\
        \hline
        Ornstein-Uhlenbeck process (1) & $2.08 \pm 0.08$ & $2.14 \pm 0.04$  & $2.21 \pm 0.02$ & $2.20 \pm 0.01$  \\
        \hline
        Logistic Map & $2.62 \pm 0.15$ & $4.19 \pm 0.29$ & $4.28 \pm 0.14$ & $4.34\pm 0.03$ \\
        \hline
        \hline
        White Noise  &$2.43 \pm 0.12$ & $3.23 \pm 0.15$ & $3.24 \pm 0.07$ & $3.32 \pm 0.02$  \\
        \hline
        Ornstein-Uhlenbeck process (2)  & $2.62 \pm 0.17$ & $2.05 \pm 0.04$ & $2.01 \pm 0.02$  & $2.00 \pm 0.01$  \\
        \hline
        Lévy flight & $2.45 \pm 0.12$ & $2.40 \pm 0.07$ &  $2.39 \pm 0.03$ & $2.37 \pm 0.01$ \\
        \hline
        \hline
        Ornstein-Uhlenbeck process (3)  & $2.34 \pm 0.11$ & $1.63\pm 0.02$ & $1.64 \pm 0.01$ & $1.60 \pm 0.01$ \\
        \hline
        Fractional Brownian Motion (2)  & $1.74 \pm 0.05$ & $1.60 \pm 0.02$ & $1.49 \pm 0.01$ & $1.49 \pm 0.01$\\
        \hline
        Pink Noise  & $3.20 \pm 0.26$ & $2.85\pm0.08$ &  $2.92 \pm 0.05$ & $2.94 \pm 0.01$  \\
        \hline
    \end{tabular}
    \caption{Estimated degree distribution exponents $\gamma$ for time series of different lengths.}
    \label{tab: gammas_lengths}
\end{table}

\begin{table}[hp]
    \centering
    \begin{tabular}{|c|c|c|c|c|}
        \hline
         \textbf{Time Series} &  $D \, (250)$ & $D \, (10^3)$& $D \, (5 \times 10^3)$ & $D \, (10^5)$  \\
         \hline
         Logistic Map & 0.17 & 0.12 & 0.15 & 0.13\\
         \hline
         Ornstein-Uhlenbeck process (1) & 0.15 & 0.13 &  0.10 & 0.10\\
         \hline
         Fractional Brownian Motion (1) & 0.12 & 0.11 & 0.08 & 0.07\\
         \hline
         Blue Noise & 0.12 & 0.08 & 0.09 & 0.08\\
         \hline
         \hline
         White Noise & 0.15 & 0.10 & 0.08 & 0.08\\
         \hline
         Ornstein-Uhlenbeck process (2) & 0.10 & 0.11 & 0.11 & 0.07\\
         \hline
         Lévy flight & 0.11 & 0.07 & 0.08 & 0.06\\
         \hline
         \hline
          Ornstein-Uhlenbeck process (3) & 0.11 & 0.14 & 0.11 & 0.09\\
         \hline
         Fractional Brownian Motion (2) & 0.21 & 0.16 & 0.17 & 0.09\\
         \hline
         Pink Noise & 0.15 & 0.09 & 0.11 & 0.08\\
         \hline
    \end{tabular}
    \caption{Kolmogorov–Smirnov distance $D$ corresponding to the fits for the various time series lengths.}
    \label{tab: KS_distance}
\end{table}

%%%%%%%%%%%%%%%%%%%%%%%%%%%%%%%%%%%%%%%%%%%%%%%%%%%%%%%%%%%%%%%%%%%%%%%%%%%%%%%%%%%%%%%
\clearpage
\subsection{Analyzing critical and Hurst exponents}

With the critical exponent $\gamma$ obtained for each time series, corresponding to the final point ($N=5000$) shown in Fig.~\ref{fig: vg_validation}, we can now classify the series according to their Hurst exponent $H$. 

%%%%%%%%%%%%%%%%%%%%%%%%%%%%%%%%%%%%%%%%%%%%%%%%%%%%%%%%%%%%%%%%%%%%%%%%%%%%%%%%%%%%%%%
\subsubsection{Time series with $0<H<0.5$}
\label{tsmenorH}

When a time series has a Hurst exponent less than 0.5, it is considered antipersistent. In this case, we have the logistic map in the chaotic regime ($r=3.8$), the Ornstein-Uhlenbeck process, a fractional Brownian motion, and blue noise. The first corresponds to a deterministic (chaotic) system, while the latter represents stochastic processes.

\begin{figure}[hp]
    \centering
    \includegraphics[width=1.0\linewidth]{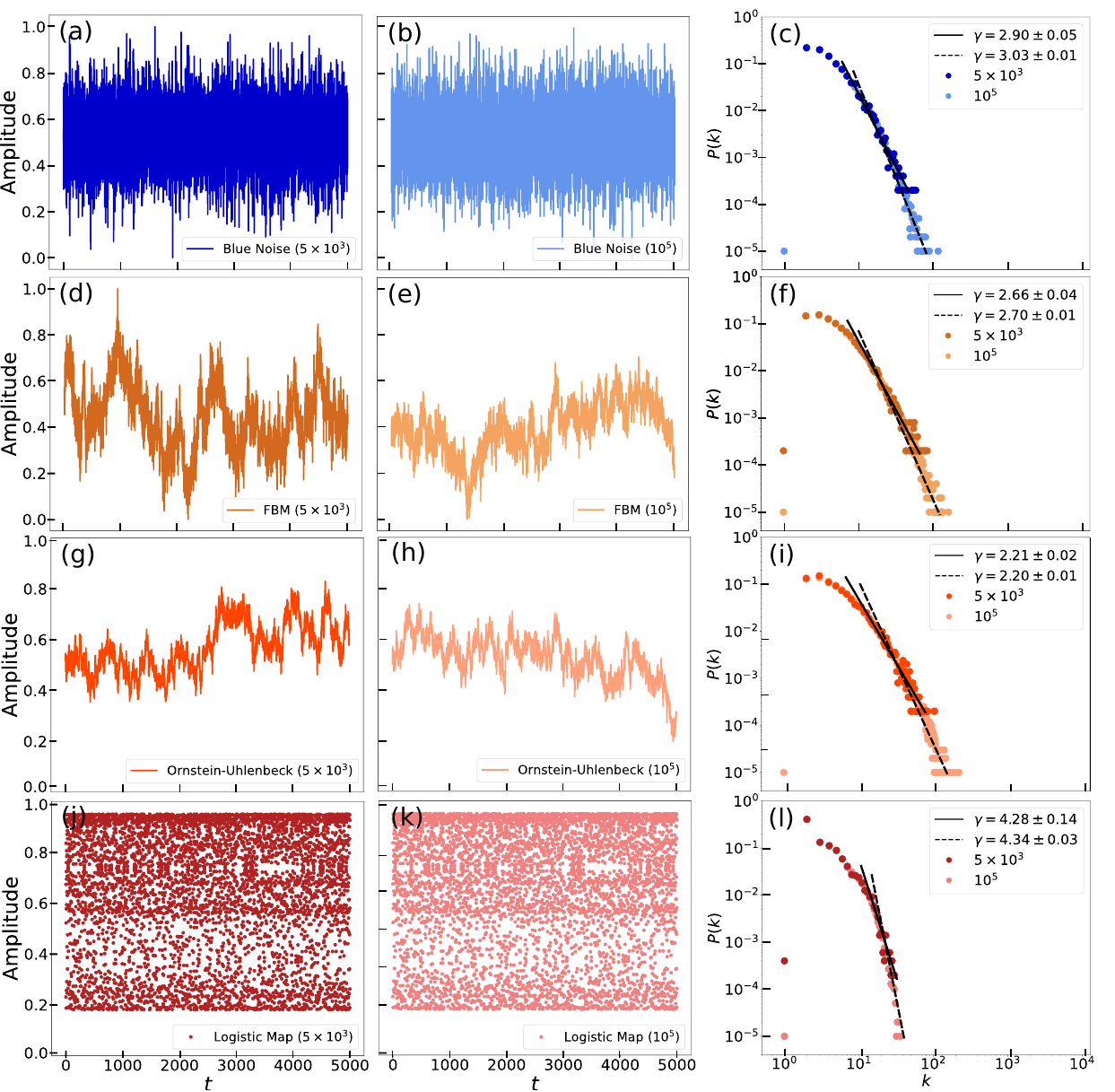}
    \caption{Results for time series with $H<0.5$. Panels (a), (d), (g), and (j) show time series of 5000 data points, while panels (b), (e), (h), and (k) display the first 5000 points of time series with total length $10^5$. Panels (c), (f), (i), and (l) present the corresponding degree distributions, with black lines indicating the MLE linear fits. Specifically: (a–c) Blue Noise with $H=0.1$ and $k_{\text{min}}=6$; (d–f) fractional Brownian motion (FBM) with $H=0.2$ and $k_{\text{min}}=7$, denoted in Fig.~\ref{fig: vg_validation} as (1); (g–i) Ornstein–Uhlenbeck process with $H=0.3$ and $k_{\text{min}}=5$, denoted in Fig.~\ref{fig: vg_validation} as (1); and (j–l) Logistic Map with $H=0.4$ and $k_{\text{min}}=8$.} 
    \label{fig: pk_h<05}
\end{figure}

In Fig.~\ref{fig: pk_h<05}, the left (5\,000 points) and center (100\,000 points) columns illustrate the temporal dynamics of the different time series analyzed, whereas the right-hand side shows the corresponding degree distributions obtained from their associated visibility graphs. The quantitative details are presented in more depth in Table~\ref{tab: table}. In this Figure, it can be observed that the logistic map does not exhibit a consistent power-law behavior. As established by Broido \& Clauset \cite{broido2019scale}, the classification of a network as strongly scale-free requires that its degree distribution follows a power-law with an exponent $\gamma$ strictly between 2 and 3. Therefore, we cannot find a power-law in this chaotic system, i.e., the logistic map (in the chaotic regime) is not scale-free. Furthermore, the other 3 time series satisfy this condition, although their behavior is quite different. The fact that the logistic map does not exhibit scale-free behavior may be beneficial. This absence suggests a lack of cross-scale interactions, meaning that the system's dynamics do not preferentially couple fluctuations across different temporal or amplitude scales. In other words, the magnitude of a given fluctuation is statistically independent of whether a large or small fluctuation preceded it. There is no memory or preference in the sequence of events. This behavior is characteristic of systems with short-range correlations, in contrast to those exhibiting long-range dependence or hierarchical organization.

%%%%%%%%%%%%%%%%%%%%%%%%%%%%%%%%%%%%%%%%%%%%%%%%%%%%%%%%%%%%%%%%%%%%%%%%%%%%%%%%%%%%%%%
\newpage
\subsubsection{Time series with $H=0.5$}

\begin{figure}[hp]
    \centering
    \includegraphics[width=1.\linewidth]{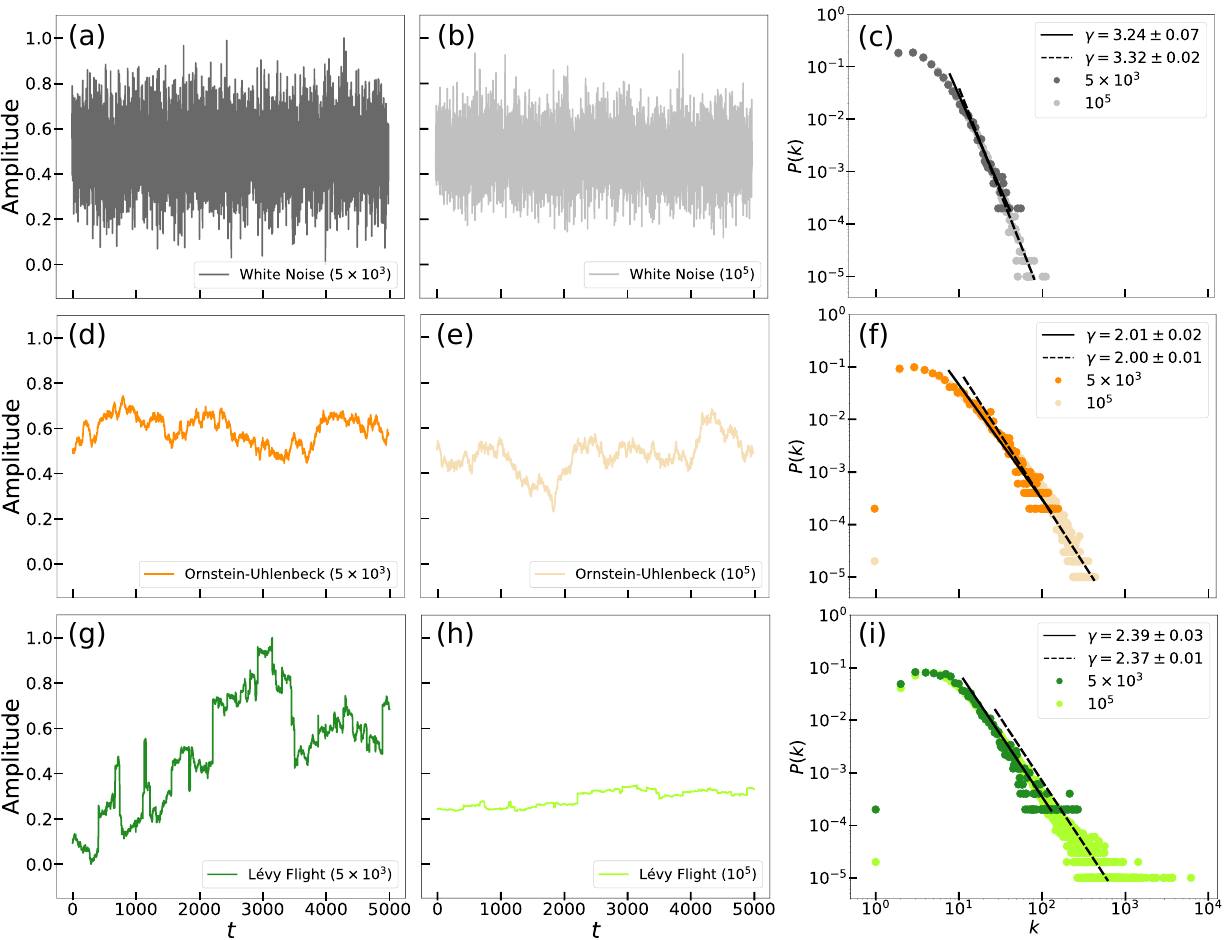}
    \caption{Results for time series with $H=0.5$. Panels (a), (d), and (g) show time series of 5000 data points, while panels (b), (e), and (h) display the first 5000 points of the time series with total length $10^5$. Panels (c), (f), and (i) present the corresponding degree distributions, with black lines indicating the MLE linear fits. Specifically: (a–c) White Noise with $k_{\text{min}}=8$; (d–f) Ornstein–Uhlenbeck process  with  $k_{\text{min}}=6$, denoted in Fig.~\ref{fig: vg_validation} as (2); and (g–i) Lévy flight with  $k_{\text{min}}=8$.} 
    \label{fig: pk_h=05}
\end{figure}

If a time series has a Hurst exponent equal to 0.5, it behaves as a completely random process. In this case, we consider three stochastic processes: white noise, the Ornstein-Uhlenbeck process, and a Lévy flight with $\alpha=1.3$, following the work of Masoomy et al.~\cite{masoomy2023PhysRevE}. This specific value of $\alpha$ is chosen because it is the value for which a Hurst exponent with value 0.5 is obtained. 
%the behavior of the critical exponent $\gamma$ is symmetric and qualitatively similar across different values of $\alpha$. 
The time series and their corresponding degree distributions are presented in Fig.~\ref{fig: pk_h=05}.

This figure highlights the markedly different behaviors among the time series: white noise appears highly irregular, the Ornstein-Uhlenbeck process exhibits monofractal characteristics, and the Lévy flight displays multifractal properties. Despite these differences, all three share the same Hurst exponent. Among them, only white noise fails to meet the criterion for being considered strongly scale-free~\cite{broido2019scale}, although the deviation is relatively minor.

%%%%%%%%%%%%%%%%%%%%%%%%%%%%%%%%%%%%%%%%%%%%%%%%%%%%%%%%%%%%%%%%%%%%%%%%%%%%%%%%%%%%%%%

\newpage
\subsubsection{Time series with $0.5<H<1$}

\begin{figure}[hp]
    \centering
    \includegraphics[width=1.\linewidth]{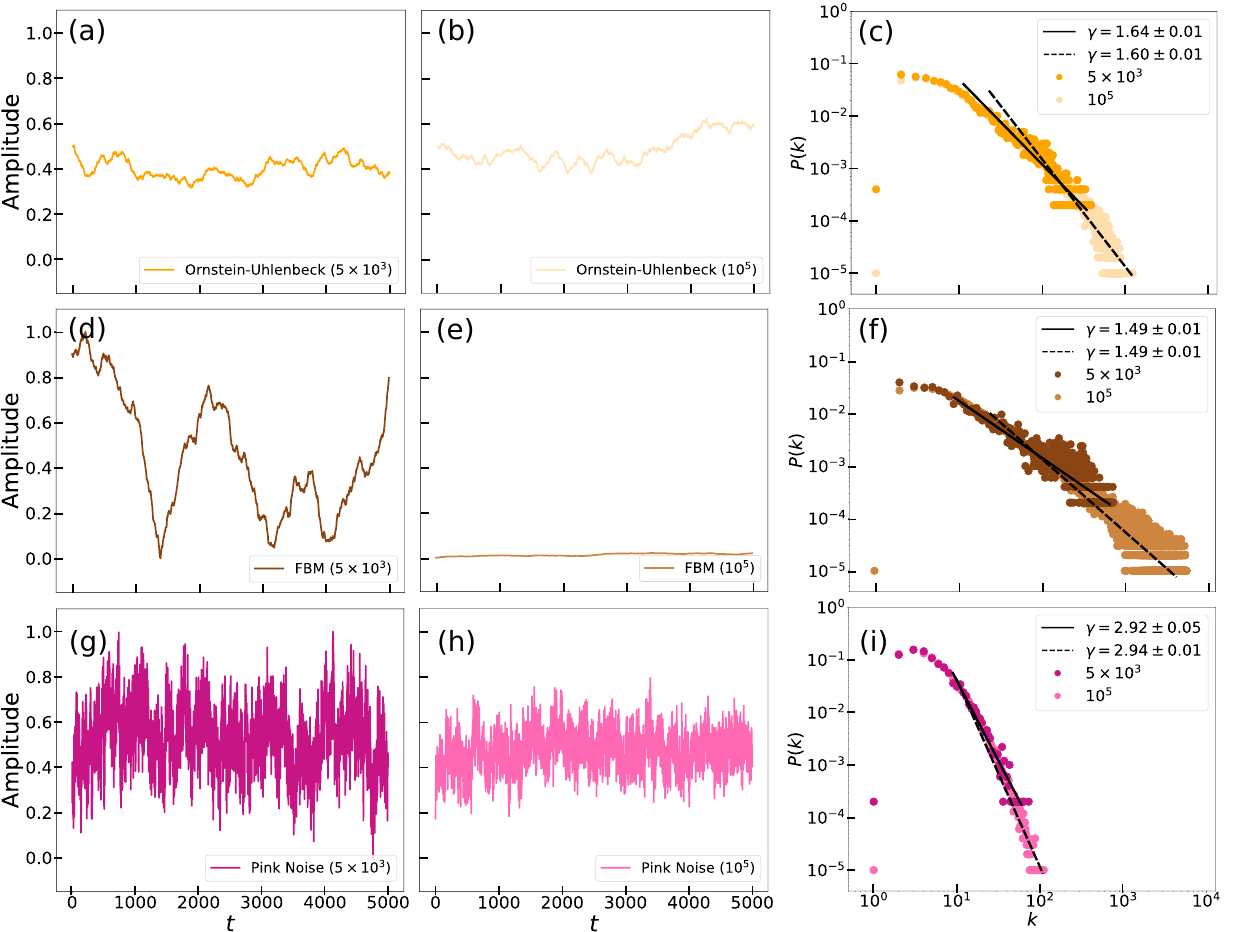}
    \caption{Results for time series with $H>0.5$. Panels (a), (d), and (g) show time series of 5000 data points, while panels (b), (e), and (h) display the first 5000 points of the time series with total length $10^5$. Panels (c), (f),  and (i) present the corresponding degree distributions, with black lines indicating the MLE linear fits. Specifically: (a–c) Ornstein-Uhlenbeck process with $H=0.7$ and $k_{\text{min}}=6$, denoted in Fig.~\ref{fig: vg_validation} as (3); (d–f) fractional Brownian motion (FBM) with $H=0.8$ and $k_{\text{min}}=7$, denoted in Fig.~\ref{fig: vg_validation} as (2); (g–i) Pink Noise with $H=1$ and $k_{\text{min}}=8$.} 
    \label{fig: pk_h>05}
\end{figure}

If a time series has a Hurst exponent bigger than ~0.5, the series is persistent. In this context, we analyze three stochastic processes: pink noise, fractional Brownian motion (fBm), and Ornstein-Uhlenbeck process. The time series and their respective degree distributions are shown in Fig.~\ref{fig: pk_h>05}.

We can notice that in this case, only the pink noise satisfies the condition to be considered a strong scale-free system. On the other hand, the fractal systems show heavy-tailed degree distributions, with $\gamma < 2$, coinciding with the one established by Broido \& Clauset \cite{broido2019scale}.

%%%%%%%%%%%%%%%%%%%%%%%%%%%%%%%%%%%%%%%%%%%%%%%%%%%%%%%%%%%%%%%%%%%%%%%%%%%%%%%%%%%%%%%
\subsection{Relation between $\gamma$ and $H$}
\label{gamma_H}
With both the critical exponent $\gamma$ and the Hurst exponent $H$ calculated for each time series analyzed in this study, we can explore the relationship between their behaviors. As a first step, the sorted results are presented in Table~\ref{tab: table}.

\begin{figure}[hp]
    \centering
    \includegraphics[width=0.9\linewidth]{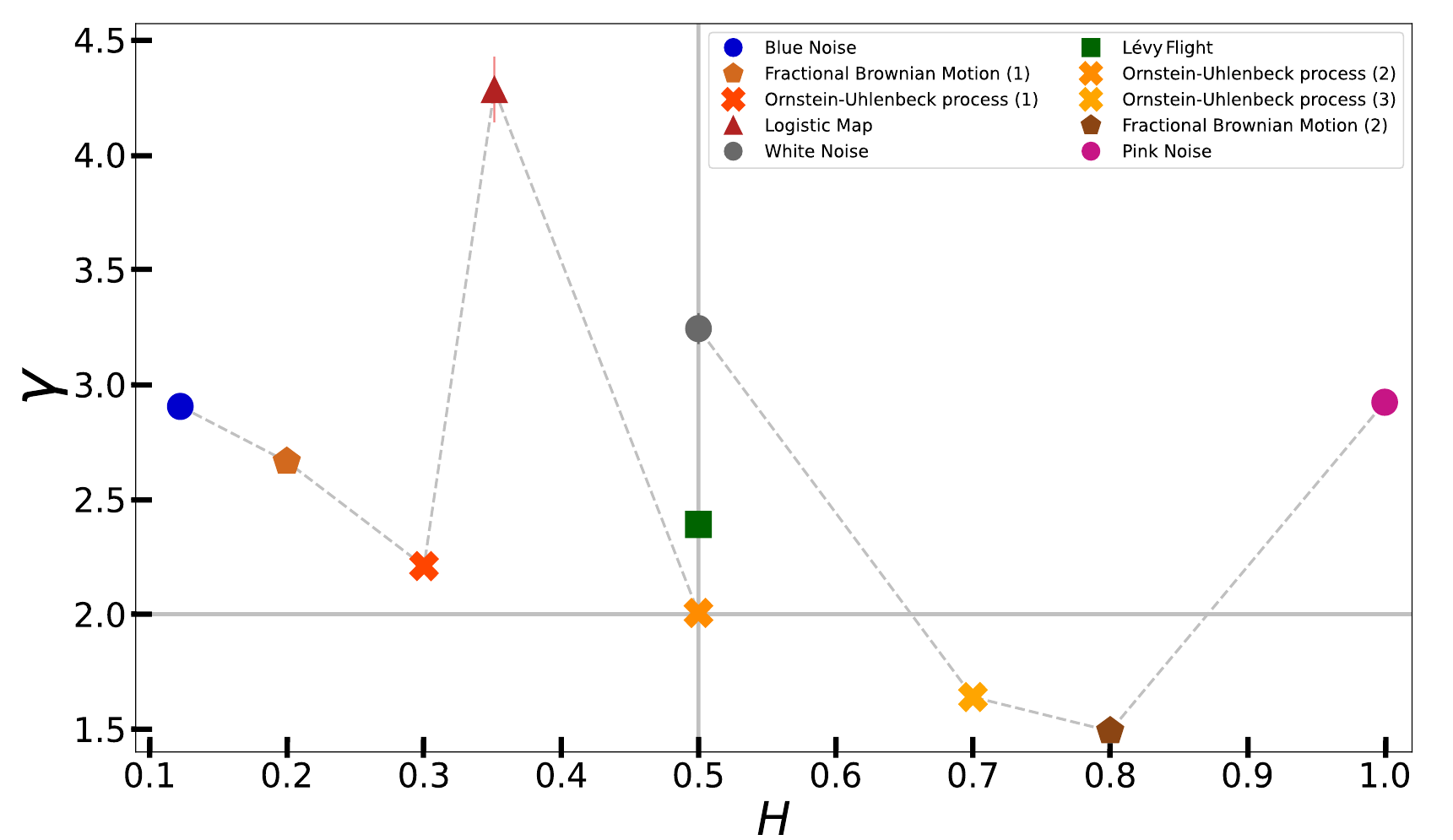}
    \caption{Comparison between $\gamma$ and Hurst exponent of each time series. The vertical gray line represents the separation between $H<0.5$ and $H>0.5$, and the horizontal line highlights the critical exponent value $\gamma=2$. Error bars for the $\gamma$ values are included in the plot; however, they are so small that they are not visible at this scale.} 
    \label{fig: gamma_hurst}
\end{figure}

Table~\ref{tab: table} reveals a noticeable correlation when $H < 0.5$, where we consistently observe $\gamma > 2$. This relationship becomes even more evident when visualized, as shown in Fig.~\ref{fig: gamma_hurst}. Interestingly, a similar trend appears for time series with $H = 0.5$, which also yields $\gamma \geq 2$ in all cases. These two findings are particularly valuable, as the Visibility Graph algorithm is relatively simple and computationally efficient to apply. Consequently, if a time series yields $\gamma \geq 2$, it is highly likely its Hurst exponent satisfies $H \leq 0.5$. The results for $H > 0.5$ are not so clear, since only in 2 of the 3 cases analyzed is there a tendency to obtain values of $\gamma < 2$. We highlight that the fBm complies with the empirical relationship obtained by Lacasa et al.~\cite{lacasa2009visibility} $\gamma(H) = 3.1 - 2\,H$
 with $N=$ 5\,000 and it is also consistent with previous results obtained in the literature~\cite{hui_2009,masoomy2023PhysRevE}. In the case of the Lévy flight with $\alpha = 1.3$ the result is also consistent with previous works~\cite{masoomy2023PhysRevE}. The different colored noises show a particular behavior, which seems to be outside the behavior of the other time series. All of them have very large $\gamma$ values, above 2.90. Regardless of the value of the Hurst exponent, that is, regardless of how persistent the time series is, the $\gamma$ values are high. Another special case seems to be the logistic map; in this study, we use the chaotic regime, and we see that the obtained value of $ \gamma$ is the highest of all the time series used, being 4.28. It is important to mention that the behavior of the $P(k)$ of the logistic map was not well adjusted by a power-law, it seems to be more of an exponential function, which would explain this high value of $\gamma$. Meanwhile, well-known processes, such as fBm and Ornstein-Uhlenbeck, show a relationship between the persistence of the time series, through the Hurst exponent, and the $\gamma$ values. Thus, we see that for antipersistent time series (0.1 $<H<$ 0.5), the values of $\gamma$ grow, are greater than 2, while for time series of the two diffusive systems for Hurst exponents associated with persistent time series (0.5 $<H<$ 1.0), the values of $\gamma$ are small and are below 2. Interestingly, all cases with $H = 0.5$ give values of $\gamma \geq 2$.

\begin{table}[h]
    \centering
    \begin{tabular}{|c|c|c|c|}
        \hline
        \textbf{Time series} & \textbf{Memory} &  $\gamma$ & $H$\\
        \hline
        Blue Noise & Non-Markovian & $2.90 \pm 0.05$ & $0.1$\\
        \hline
        Fractional Brownian Motion (1) & Non-Markovian & $2.66 \pm 0.04$ & $0.2$\\
        \hline
        Ornstein-Uhlenbeck process (1) & Markovian & $2.21 \pm 0.02$ & $0.3$\\
        \hline
        Logistic Map & Markovian & $4.28 \pm 0.14$ & $0.4$ \\
        \hline
        \hline
        White Noise & Non-Markovian & $3.24 \pm 0.07$ & $0.5$\\
        \hline
        Ornstein-Uhlenbeck process (2) & Markovian & $2.01 \pm 0.02$ & $0.5$\\
        \hline
        Lévy flight  & Markovian & $2.39 \pm 0.03$  & $0.5$\\
        \hline
        \hline
        Ornstein-Uhlenbeck process (3) & Markovian & $1.64 \pm 0.01$ & $0.7$\\
        \hline
        Fractional Brownian Motion (2) & Non-Markovian & $1.49 \pm 0.01$ & $0.8$\\
        \hline
        Pink Noise & Non-Markovian & $2.92 \pm 0.05$ & $1.0$\\
        \hline
    
    \end{tabular}
    \caption{Classification of the time series with 5000 data points studied in this work, with their respective memory and the critical and Hurst exponents. Errors in the $H$ values were computed but are negligible due to their minimal magnitude.}
    \label{tab: table}
\end{table}

%%%%%%%%%%%%%%%%%%%%%%%%%%%%%%%%%%%%%%%%%%%%%%%%%%%%%%%%%%%%%%%%%%%%%%%%%%%%%%%%%%%%%%%
% Discussion 
%%%%%%%%%%%%%%%%%%%%%%%%%%%%%%%%%%%%%%%%%%%%%%%%%%%%%%%%%%%%%%%%%%%%%%%%%%%%%%%%%%%%%%%

\newpage
\subsection{Comparative Analysis of Slope Exponents: VG vs. HVG}

While our analysis is based on the standard Visibility Graph algorithm, it is useful to contextualize our findings by referring to the closely related Horizontal Visibility Graph (HVG), which has been extensively studied in the literature~\cite{xie_2011, Yu_2016, gonccalves2016time, braga2016characterization, acosta2021applying, baoli2023, azizi2024review}. The HVG framework has been widely applied to time series of various types, and its properties are well characterized both numerically and theoretically. In particular, a key and well-established result is that the degree distribution, $P(k)$, of HVG networks usually decays exponentially as $P(k) \sim \exp(-\lambda k)$, where $\lambda$ acts as a slope parameter reflecting the degree of correlation or randomness in the original time series. Given the depth of insight provided by HVG-based analyses, drawing parallels with these results helps to better interpret the behavior observed in our VG-based approach.

One of the classical results in HVG analysis, reported by Luque et al.~\cite{luque2009horizontal}, is that the degree distribution associated with uncorrelated random processes (white noise) exhibits a universal exponential decay with a slope $\lambda = \ln(3/2) \approx 0.405$. This value is often taken as a useful reference point to interpret the behavior of $\lambda$ across different types of time series, providing a baseline from which deviations can be understood in terms of the underlying temporal structure. In particular, time series with inherent correlations, such as those generated by autoregressive models, chaotic maps, or deterministic dynamics, typically yield slower decay rates, i.e., smaller $\lambda$ values. These lower slopes reflect the increased regularity or predictability embedded in the signal. In light of this, it becomes particularly relevant to examine whether the slopes obtained via our VG-based analysis follow similar trends when compared to those well-established HVG benchmarks.

Unlike HVG, where the degree distribution typically follows an exponential decay, the standard VG method often results in a power-law decay, as shown in Eq.~\eqref{eq: power_law}. As a consequence, the exponents $\lambda$ and $\gamma$ are not directly comparable in a quantitative sense. However, both can be interpreted as measures of structural complexity or correlation: lower values of $\lambda$ or $\gamma$ are generally associated with more regular or correlated dynamics. This conceptual parallel allows for a qualitative comparison between our VG-derived results and existing findings from the HVG literature.

Although the slope parameters $\lambda$ and $\gamma$ characterize different functional forms, our results suggest that both respond similarly to the presence of temporal correlations in the underlying time series. In particular, time series with either persistent ($H > 0.5$) or antipersistent ($H < 0.5$) structure tend to deviate from the white noise benchmark in both frameworks. For HVG, this typically manifests as a reduction in the slope $\lambda$ below the reference value $\ln(3/2) \approx 0.405$, as shown in \cite{gonccalves2016time, ravetti2014distinguishing}. Similarly, in our VG-based analysis, correlated processes yield lower values of $\gamma$ compared to the maximum value observed for white noise ($\gamma \approx 3.24$), indicating heavier tails and increased graph connectivity. In this analysis, we have excluded the logistic map time series, which has a degree distribution that does not decay as a power-law and actually appears to exhibit an exponential decay, as mentioned earlier. Notably, both approaches appear to capture the presence of correlation but not necessarily its direction: HVG-based studies report that persistent and antipersistent series with the same distance from $H = 0.5$ often produce comparable $\lambda$ values. This asymmetry in sensitivity suggests that $\gamma$, like $\lambda$, may serve as a robust but coarse indicator of correlation strength rather than sign.

Despite the interpretive value of slope-based parameters such as $\lambda$ in HVG and $\gamma$ in VG, these single-metric approaches may fall short in fully capturing the diverse range of dynamical behaviors present in complex time series. In particular, neither exponent is capable of distinguishing between persistent and antipersistent correlations with high precision, as both tend to respond symmetrically to the distance from the uncorrelated case ($H = 0.5$). This limitation highlights the necessity of incorporating additional descriptors to obtain a more complete understanding of temporal structure. For example, previous studies have successfully combined graph-based measures with information-theoretic quantifiers, such as Shannon entropy and Fisher information \cite{ravetti2014distinguishing}, or statistical complexity measures derived from ordinal patterns \cite{rosso2007distinguishing}. These multidimensional approaches not only enhance the discriminatory power between stochastic and deterministic dynamics, but also allow for a more nuanced assessment of structure, randomness, and memory effects within the system. In this light, our findings on $\gamma$ should be regarded as a foundational yet complementary piece within a broader analytical framework.

%%%%%%%%%%%%%%%%%%%%%%%%%%%%%%%%%%%%%%%%%%%%%%%%%%%%%%%%%%%%%%%%%%%%%%%%%%%%%%%%%%%%%%%
% Conclusions
%%%%%%%%%%%%%%%%%%%%%%%%%%%%%%%%%%%%%%%%%%%%%%%%%%%%%%%%%%%%%%%%%%%%%%%%%%%%%%%%%%%%%%%
\section{Conclusions}
\label{sec: conclusions}

We have explored how structural features extracted from the Visibility Graph relate to the temporal properties of various time series, with a particular focus on the role of memory and persistence. By examining the degree distribution exponent $\gamma$ obtained from the VG and comparing it with the well-established Hurst exponent $H$, we sought to uncover potential links between network topology and time series dynamics. While the Hurst exponent is widely used to quantify temporal correlations, the exponent $\gamma$ remains relatively underexplored, despite its promise as a complementary measure capturing cross-scale structural information. In this work, we analyzed how $\gamma$ behaves across a range of systems characterized by distinct memory properties, spanning deterministic, stochastic, Markovian, and non-Markovian dynamics. Our investigation proceeds in three main stages:

First, we evaluated the sensitivity of the VG algorithm to ensure the reliability of the resulting $\gamma$ values. We found that, for consistent and confident use of the VG method, a minimum of 1000 data points is recommended. Nonetheless, meaningful results can still be obtained with as few as 500 data points, provided appropriate caution is exercised.

Second, we analyzed the degree distribution by applying the VG algorithm to 10 different time series. These time series were classified according to their Hurst exponents. Our results show that the logistic map does not exhibit scale-free behavior, as its exponent $\gamma$ lies well outside the expected range for strong power-law behavior ($2< \gamma <3$). Similarly, white noise does not satisfy the conditions for scale-freeness. These findings suggest that not all systems exhibit scale-free behavior, regardless of whether their dynamics are chaotic or stochastic. The presence of such behavior appears to be more nuanced and dependent on the underlying memory and correlation structure of the system.

Third, we investigated the relationship between the critical exponent $\gamma$ and the Hurst exponent $H$ for each time series. Our results reveal that when $H \leq 0.5$, the corresponding time series consistently exhibit a critical exponent $\gamma \geq 2$. This suggests a lower bound on $\gamma$ linked to persistent or memory-retaining dynamics, reinforcing the connection between long-range correlations in the time series and the emergence of heavy-tailed degree distributions in their visibility graphs. However, when $H > 0.5$, the behavior of the critical exponent becomes less predictable, and no consistent pattern can be established with confidence. These are general results when observing all values of $H$ and $\gamma$, however, the results for each time series contain other details. The time series of the logistic map does not show a power-law, so the exponent $\gamma$ has a very high value, which indicates that using the VG is not the appropriate method to study this system, unlike the HVG, which has shown interesting results when studying the logistic map~\cite{Luque_pone2011}. On the other hand, the different colored noises do not seem to have a direct relationship between the Hurst exponent and the $\gamma$ exponent. About Lévy flight, we obtain a value of $\gamma > 2.0$ for the case studied. This is consistent with all series with $H \leq 0.5$. The series that do show a linear relationship between the Hurst exponent and the $\gamma$ exponent are the fBm and the Ornstein-Uhlenbeck process. Both model diffusive processes and, are the only time series that do show linear behavior with $H$. Thus, for these two series, we can conclude that if the process is diffusive and antipersistent, the values of $\gamma$ will be greater than 2.0; if the process is random, the value of $\gamma$ will be close to 2.0; and if the process is persistent, the value of $\gamma$ will be less than 2.0. The details of these behaviors will be addressed in another work.

Finally, it is not possible to conclude that there is a direct relationship between the values of the decay $\lambda$ in the case of HVG and the exponent of the power-law of $\gamma$ of the VG; however, a possible relationship between both exponents could be explored in future research.

In summary, this study opens up new possibilities for studying different time series through the VG method and, not only the HVG method, which has been widely used to analyze different time series.
We highlight a meaningful connection between the critical exponent $\gamma$ and the Hurst exponent $\gamma$ for time series that model diffusive processes, offering a complementary perspective on the memory and correlation properties inherent to time series data of diffusive processes. These results not only deepen our understanding of how complex network metrics relate to classical statistical measures but also establish practical guidelines for applying the VG algorithm to real-world time series with varying lengths and dynamical behaviors.

%%%%%%%%%%%%%%%%%%%%%%%%%%%%%%%%%%%%%%%%%%%%%%%%%%%%%%%%%%%%%%%%%%%%%%%%%%%%%%%%%%%%%%%
% Acknowledgements
%%%%%%%%%%%%%%%%%%%%%%%%%%%%%%%%%%%%%%%%%%%%%%%%%%%%%%%%%%%%%%%%%%%%%%%%%%%%%%%%%%%%%%%
\section*{Acknowledgments}

We thank the support from ANID Chile through the MSc grant No. 22240809 (M.C.) and the FONDECYT grant No. 1240281 (P.S.M.).
I.G.M. is grateful for the support from a postdoctoral project No. 0016500-2025 funded by the University of Chile, which financed his participation in this work. 

\bibliographystyle{elsarticle-num}
\bibliography{references} 

\end{document}